\documentclass[preprint,showpacs,aps,preprintnumbers,amsmath,amssymb,
nofootinbib]{revtex4}
\usepackage{epsfig}
\begin{document}

\begin{flushright}
\end{flushright}


\newcommand{\be}{\begin{equation}}
\newcommand{\ee}{\end{equation}}
\newcommand{\bea}{\begin{eqnarray}}
\newcommand{\eea}{\end{eqnarray}}
\newcommand{\bers}{\begin{eqnarray*}}
\newcommand{\eers}{\end{eqnarray*}}
\newcommand{\nn}{\nonumber}

\def\lp{\lambda^\prime}
\def\ll{\Lambda}
\def\mb{m_{\Lambda_b}}
\def\ml{m_\Lambda}
\def\s1{\hat s}
\def\ds{\displaystyle}


\title{\large Probing new physics in $B \to f_0(980) K$ decays  }
\author{A. K. Giri$^1$, B. Mawlong$^2$ and
R. Mohanta$^2$ }
\affiliation{$^1$ Department of Physics, Punjabi University,
Patiala - 147002, India\\
$^2$ School of Physics, University of Hyderabad, Hyderabad - 500 046,
India}


\begin{abstract}
We study the hadronic decay modes $B^{ \pm(0)} \to 
f_0(980) K^{\pm(0)} $, involving a scalar and 
a pseudoscalar meson in the final state.
These decay modes are dominated by the loop induced $ b \to s \bar q q~
(q=s,~u,~d)$ 
penguins along with a small $ b \to u $ tree level 
transition (for $B^+ \to f_0 K^+$) 
and  annihilation  diagrams.
Therefore, the standard model expectation of direct CP violation is 
negligibly small and the mixing induced CP violation parameter
in the mode $B^0 \to f_0 K_S$ is expected to give the same value
of $\sin(2 \beta)$, 
as extracted from $B^0 \to J/\psi K_S$ but with opposite sign. 
Using the generalized factorization approach we find the direct 
CP violation in the decay mode $B^+ \to f_0 K^+ $ to be 
of the order of few percent.
We then study the effect of
the R-parity violating supersymmetric model  and show that the 
direct CP violating
asymmetry in $B^+ \to f_0(980) K^+ $  could be as large as $\sim 80 \% $
and the mixing induced CP asymmetry in 
$B^0 \to f_0 K_S$ (i.e., $-S_{f_0 K_S}$)  could
deviate significantly from that of $\sin(2 \beta)_{J/\psi K_S}$.
\end{abstract}

\pacs{13.25.Hw, 11.30.Er, 12.60.Jv}
\maketitle

\section{Introduction}
The currently running $B$ factories, such as Belle and Babar, 
are providing us huge data in the b-quark sector. The main objectives
of these B factories is to critically test the standard model (SM)
predictions and to look for possible signature of new physics.
For this purpose, a variety of useful observables are being measured
and are compared with the corresponding theoretical predictions.
One of the important observables of this kind is the CP asymmetry parameter
in various $B$ meson decays and the other one being the branching ratio for
rare $B$ processes.

Recently, both Belle \cite{belle1,belle2,bel3,bel4} and Babar 
\cite{bar1,bar2,bar3,babar3,babar4}
have reported the measurement of branching ratios and
CP violating parameters in the rare decay modes $B^{0,+} \to f_0(980) 
K^{0,+}$, involving a scalar and a pseudoscalar meson in the final state.
The measured decay rates for the mode $B^+ \to f_0 K^+ $ are
\bea
BR(B^+ \to f_0(980) K^+ \to \pi^+ \pi^- K^+) &=& 
(8.78 \pm 0.82_{-1.76}^{+0.85})
\times 10^{-6}\;,~~~~~~[3]\nn\\ 
BR(B^+ \to f_0(980) K^+ \to \pi^+ \pi^- K^+) &=& (9.47 \pm 0.97_
{-0.88}^{+0.62} )
\times 10^{-6}\;,~~~~~~[8]
\eea
with an average 
\be
BR(B^+ \to f_0(980) K^+ \to \pi^+ \pi^- K^+) = (9.21 \pm 0.97)
\times 10^{-6}\;.
\ee
For the process $B^0 \to f_0 K^0 $, the measured rates are
\bea 
BR(B^0 \to f_0(980) K^0 \to \pi^+ \pi^- K^0)& = &(7.60 \pm 1.66
_{-0.89}^{+0.76})
\times 10^{-6}\;,~~~~~~[4]\nn\\
BR(B^0 \to f_0(980) K^0 \to \pi^+ \pi^- K^0) &=& (5.5 \pm 0.7
\pm 0.7
\times 10^{-6}\;.~~~~~~~~~[9]
\eea
The absolute branching ratios for $B \to f_0 K$ processes
depend on the branching fraction of $f_0 \to \pi^+ \pi^-$ 
process. Using the results from \cite{ani} for
$\Gamma(f_0 \to \pi \pi)=64 \pm 8 $
MeV,
$\Gamma_{f_0}^{tot}=80 \pm 10 $ MeV
along with the relation
$\Gamma(f_0 \to \pi^+ \pi^-)=\frac{2}{3}\Gamma(f_0 \to \pi \pi)$,
 we obtain
the branching ratios  for $B \to f_0 K$ processes as
\bea
BR(B^+ \to f_0(980) K^+ )& = & (17.38 \pm 3.47)
\times 10^{-6}\;, \nn\\
BR(B^0 \to f_0(980) K^0 ) & = &(11.26 \pm 
2.52)
\times 10^{-6}\;.
\eea

The mixing induced parameter for the process $B^0 \to f_0 K_S$, 
observed by both  Babar and Belle as 
\bea
\sin (2 \beta)_{f_0 K_S} &=& 0.95_{-0.32}^{+0.23} \pm 0.10\;,
~~~~~~~~~~~~~~~~[10]\nn\\
\sin (2 \beta)_{f_0 K_S} &=& 0.18\pm 0.23\pm 0.11 \;,
~~~~~~~~~~~~[11]
\eea
with an average
\bea
\sin (2 \beta)_{f_0 K_S} = 0.51\pm 0.19\;,
\eea
which has nearly one sigma deviation from that of 
$\sin (2 \beta)_{b \to c \bar c s} 
= 0.687\pm 0.032$ \cite{hfag}. 
These observations not only provide us another way to test the 
SM and/or to look for new physics but also may help us to understand 
the nature of the light scalar meson $f_0(980)$. It should be noted 
here that the mixing induced CP violation
parameter seems to be, at present, 
not deviated significantly from its SM expectation.
But, since  the error 
bars are quite large the situation is still very much conducive 
to explore some non-standard physics.

The light scalar mesons with masses below 1 GeV is considered  
as a controversial issue for a long
time. Even today, there exists no consensus on the nature of
the $f_0(980)$ and $a_0(980)$ mesons.
While the low-energy hadron phenomenology has been successfully understood
in terms of the constituent quark model, the scalar mesons are still
puzzling and the quark composition of the light scalar mesons are 
not understood with certainty. The structure of the scalar meson $f_0(980)$ 
has been discussed
for decades and appears to be still not clear. There were attempts to 
interpret
it as $K \bar K$ molecular states \cite{wein}, four quark states
\cite{jaff} and normal $q \bar q$ states \cite{torn}. However, recent
studies of $\phi \to \gamma f_0$ ($f_0 \to \gamma \gamma$) 
\cite{fazi,ani}
and $D_s^+ \to f_0 \pi^+ $ decays \cite{akle} favor the $q \bar q$ model.
Since $f_0(980)$ is produced copiously in $D_s$ decays, this supports the
picture of large $s \bar s$ component in its wave function, as the dominant
mechanism in the $D_s$ decay is $c \to s$ transition. The
prominent $s \bar s$ nature of $f_0(980)$ has been supported by
the radiative decay $ \phi \to f_0(980) \gamma$
\cite{aul}. In this interpretation,  the flavor content of $f_0$
is given by $f_0= n \bar n \sin \theta + s \bar s \cos \theta $
with $n \bar n = (u \bar u+d \bar d)/\sqrt 2$. A mixing angle
of $\theta=138^\circ \pm 6^\circ$ has been experimentally
determined from $\phi \to \gamma f_0$ decays \cite{ani}. We will follow
this structure for our study in this paper.

Theoretically, these decay modes have been studied in the standard
model using perturbative QCD \cite{chen} and QCD factorization approach
\cite{cheng, cheng2}. In this paper, 
we would like to study the decay modes $B^0 \to 
f_0(980) K^0 $  and $B^+ \to f_0(980) K^+$ using the 
generalized factorization approach. We consider
$f_0(980)$ to be composed of $f_0(980)=
n \bar n \sin \theta+s \bar s \cos \theta $ with dominant 
$s \bar s$ composition.
Therefore, these processes may be considered, at the leading order,
as dominated by $b \to s \bar s s $ penguin amplitudes. Hence, the mixing
induced CP violation in the decay mode $B^0 \to f_0 K_S$ is expected to give
the same value of $\sin(2 \beta)$ as extracted from $B^0 \to J/\psi K_S$,
with an uncertainty of $5\%$. Comparison of these two values, therefore,
could be a sensitive probe for physics beyond the SM. Since the predicted 
branching ratios available from previous studies \cite{chen, cheng, cheng2} 
are not in agreement with the 
experimental values, we would like
to see the effect of R-parity violating supersymmetric (RPV)
model  in these modes.
Moreover, since in this paper we are interested to see whether it is
possible to extract any signature of new physics (NP) from these modes
or not, 
we resort ourselves to generalized factorization approach
in analyzing these modes.

The paper is organized as follows. In section II, we
analyze these modes in the standard model. The basic formula for
CP violating parameters are presented in section III.
The contributions arising from R-parity violating model
are presented in section IV and section V contains our
conclusion.

\section{Standard model contribution}
The effective Hamiltonian describing the charmless hadronic $B$ decays
is given as 
\be 
H_{eff}=
\frac{G_F}{\sqrt{2}}\biggr[V_{ub}V_{us}^*\sum_{i=1}^2 C_i O_i-
V_{tb}V_{ts}^* \sum_{j=3}^{10}C_j O_j
 \biggr],\label{ham}
\ee
where $G_F$ is the Fermi coupling constant, $C_i$'s are the Wilson 
coefficients,
$O_{1,2}$ are the tree operators and $O_{3-10}$ are QCD and
electroweak penguin operators. 

To calculate the branching ratios of the $B \to f_0 K$ decay processes,
we adopt the generalized factorization framework to evaluate the 
hadronic matrix elements i.e., $\langle O_i \rangle= \langle
f_0 K | O_i| B \rangle $. In this approximation, 
these hadronic matrix
elements  can be parametrized in terms of the decay
constants and the form factors which are defined as
\be
\langle 0| A^\mu | K(k) \rangle = i f_K k^\mu\;,
~~~~~~ \langle 0|\bar q q  | f_0 \rangle = m_{f_0} \bar f_{f_0}^q\;,
\ee
\bea
\langle K(k) |(V-A)_\mu|B(P)\rangle & = & 
\left [ (P+k)_\mu- \left ( \frac{m_B^2-m_K^2}{q^2}\right )q_\mu
\right ] F_1^{BK}(q^2)\nn\\
&+ &\left ( \frac{m_B^2-m_K^2}{q^2}\right )q_\mu
F_0^{BK}(q^2)\;,
\eea
\bea
\langle f_0(q) |(V-A)_\mu|B(P)\rangle & = & 
i \biggr\{\left [ (P+q)_\mu- \left ( \frac{m_B^2-m_{f_0}^2}{k^2}\right )k_\mu
\right ] F_1^{Bf_0}(k^2)\nn\\
&+ &\left ( \frac{m_B^2-m_{f_0}^2}{k^2}\right )k_\mu
F_0^{Bf_0}(k^2)\biggr\}\;,\label{amp}
\eea
where $V$ and $A$ denote the vector and axial-vector currents,
 $f_K$ and $\bar f_{f_0}$ are the decay constants of
$K$ and $f_0$ mesons, $F_{0,1}(q^2)$ are the
form factors and $P,q,k$ are the momenta of $B$, $f_0$ and $K$ mesons
satisfying the relation  $q=P-k$. 

Now let us first consider the process $ B^+ 
\to f_0 K^+$. Within the SM it receives contribution 
from $ b \to u $ tree,  $b \to s \bar q q  $ (with $q=u,s$)
penguins and annihilation diagrams. 
Using Eqs. (\ref{ham})-(\ref{amp}), one
can  obtain the amplitude  in the SM as
\bea
{\cal A}(B^+ \to f_0 K^+) &=& - \frac{G_F}{\sqrt{2}}
\biggr\{\Big[V_{ub}^* V_{us} a_1-V_{tb}^* V_{ts}(a_4+a_{10}
-r_\chi(a_6+a_8))\Big] X \nn\\
&-&V_{tb}^*V_{ts}(2 a_6 -a_8) Y
-\Big[V_{ub}^* V_{us} a_1\nn\\
&-& V_{tb}^*V_{ts}\Big(a_4+a_{10}
-\frac{2(a_6+a_8) m_B^2}{(m_b+m_u)(m_s+m_u)}\Big)\Big] Z \biggr\} \;,
\eea
where
\bea
&&r_\chi=\frac{2 m_K^2}{(m_b+m_u)(m_s+m_u)}\;,~~~~~~~~~~~~~~~~
X=f_K (m_B^2-m_{f_0}^2)F_0^{Bf_0}(m_K^2)\;,\nn\\
&&Y= \bar f_{f_0}^s~ m_{f_0}\frac{m_B^2-m_K^2}{m_b -m_s}F_0^{BK}(m_{f_o}^2)\;,
~~~~~~~~
Z=f_B (m_{f_0}^2-m_K^2)F_0^{f_0 K}(m_B^2)\;,
\eea
and $a_i$'s are the combinations of Wilson coefficients given by
\bea
a_{2i-1}=C_{2i-1}+\frac{1}{N_C}C_{2i}\;,~~~
a_{2i}=C_{2i}+\frac{1}{N_C}C_{2i-1}\;,~~~~(i=1,2,3,4,5)
\eea
with $N_C$ as the number of colors.

The corresponding neutral process $B^0 \to f_0 K^0$
receives contribution only from $ b \to s \bar q q $ (with $q=s,d$)
penguins and annihilation diagrams. Thus, one can write the
amplitude\footnote{The sign of the coefficients of $Y$ in Eqs. (11) and (14)
are found to be opposite to that of Ref. [28].} for this process  as
\bea
{\cal A}(B^0 &\to& f_0 K^0) =  \frac{G_F}{\sqrt{2}}V_{tb}^*V_{ts}
\biggr\{\Big[a_4-\frac{a_{10}}{2}
-r_{\chi_1}(a_6-\frac{a_8}{2})\Big] X \nn\\
&+&(2 a_6 -a_8) Y
-\Big[ a_4-\frac{a_{10}}{2}
-\frac{(2a_6-a_8) m_B^2}{(m_b+m_d)(m_s+m_d)}\Big] Z \biggr\}\;, 
\eea
where $r_{\chi_1}$ can be obtained from $r_\chi$ by 
replacing the $K^+$ and $u$-quark masses by
$K^0$ and $d$ masses.
The  branching ratios can be obtained from these amplitudes 
as
\be
BR(B \to f_0 K)= \frac{|{\rm p_{c}}| \tau_B}{8 \pi m_B^2}|{\cal A}
(B \to f_0 K)|^2\;,
\ee
where $|{\rm p_{c}}|$ is the c.m. momentum of the final mesons
and $\tau_B$ is the lifetime of the $B$ meson.
For numerical analysis, we use the particle masses and lifetimes
from \cite{pdg}. The current quark masses 
are taken as $m_b=4.88$ GeV, $m_s=122$ MeV, $m_d=7.6$ MeV
and $m_u=4.2$ MeV.
The values of the effective QCD parameters ($a_i$'s) are taken from
\cite{ali}, which
are evaluated at the scale $\mu= m_b/2$. For the CKM matrix
elements, we use the Wolfenstein parametrization with the 
parameters $A$=0.801,
$\lambda=0.2265$, $\bar \rho=0.189$ and $\bar \eta=0.358$ \cite{ckm}.
The form
factors describing the transition $B \to f_0$ 
are given as \cite{cheng}
\bea
F_0^{B^- f_0 }= \frac{1}{\sqrt 2} \sin \theta F_0^{B^- f_0^{u\bar u}}\;,
~~~~~~~
 F_0^{B^0 f_0 }= \frac{1}{\sqrt 2} \sin \theta F_0^{B^0 f_0^{d\bar d}}\;,
\eea
with $ F_0^{B f_0^{q\bar q}}(0)$ ( $q \bar q= u \bar u $ or $d \bar d $)
being of the order of 0.25 \cite{cheng1}. For the $ q^2$
dependence, we assume the simple pole dominance as
\be
F_0^{B f_0}(q^2)= \frac{F_0^{B f_0}(0)}{1-q^2/m_P^2}\;,
\ee
with $m_P $ being the mass of the $0^-$ pole state with the same quark
content as the current under consideration. For the form factors, describing
$B \to K $ transition, we use the corresponding QCD sum rule value 
\cite{ball}
\bea
F_0^{BK}(m_{f_0}^2)=\frac{0.3302}{1-\frac{m_{f_0}^2}{37.46}}\;.
\eea
The annihilation form factor $F_0^{f_0 K}(q^2)$ is expected to be suppressed 
at large momentum transfer (i.e., $q^2=m_B^2$) due to helicity
suppression. However, it may receive long distance contributions from
nearby resonances via final state interactions. In Ref. \cite{epjc},
its value is extracted using the experimental values of $BR (B \to f_0 K)$,
where it has been shown that in order to explain the observed data
in the SM one requires large value of annihilation form factor, if
the $B \to f_0$ form factor will be $F_0^{B f_0 }\leq 0.2$. Since,
we are interested to look for new physics signature 
in this mode, here we use the
lowest value of $|F_0^{f_0 K}|$, which is around 0.03, as seen from
figure-2 of \cite{epjc}. Furthermore, since both the components 
of $f_0$ ($n \bar n$ and $s \bar s$) are
involved in the annihilation topology, the corresponding
amplitude should be multiplied by $(\sin \theta/\sqrt 2+
\cos \theta)$.  
 
The decay constants used are $f_K$=0.16 GeV, $f_B$=0.19 GeV and $\bar f_0^s=
\frac{m_{f_0}^{(s)}}{m_{f_0}} \tilde f_s \cos \theta $ with $ \tilde f_s
(\mu = 2.1 {\rm GeV})$=0.39 GeV and $m_{f_0}^{(s)} \simeq
(1.02 \pm .05)$ GeV \cite{cheng}.

Using these values and the mixing angle $\theta=138^\circ$,
we obtain the branching ratios for the $B \to f_0(980) K$ processes as
\bea
&&BR(B^+ \to f_0(980) K^+)= 6.56 \times 10^{-6}\;,\nn\\
&&BR(B^0 \to f_0(980) K^0)= 4.73 \times 10^{-6}\;,
\eea
which are quite below the experimental values (4).
The variation of the branching ratios for the strange,
non-strange mixing angle 
$\theta$ between $0$ and $\pi$ are shown in figures
1 and 2. Thus, one can see from the figure-1
that for $B^+ \to f_0 K^+$ process
generalized factorization approach cannot
accommodate the experimental data for any value of the mixing angle
$\theta $. For the $B^0 \to f_0 K^0$ mode also it 
cannot explain the data unless $\theta $ is very close to 0 or $\pi$
as seen from figure-2.

\begin{figure}[htb]
   \centerline{\epsfysize 2.0 truein \epsfbox{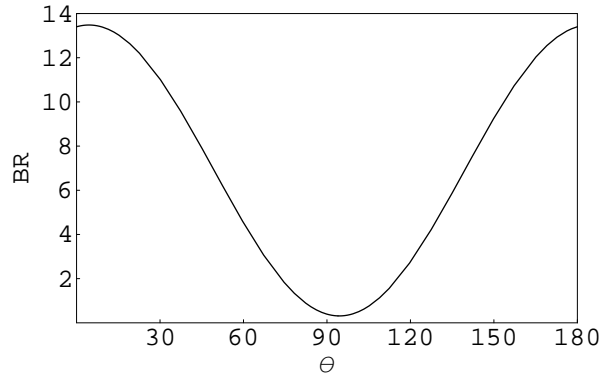}}
 \caption{
  The  branching ratio  for the process  
$B^- \to   f_0(980) K^- $ (in units of $10^{-6}$), versus the mixing 
angle $\theta$ in degrees.}
  \end{figure}
\begin{figure}[htb]
   \centerline{\epsfysize 2.0 truein \epsfbox{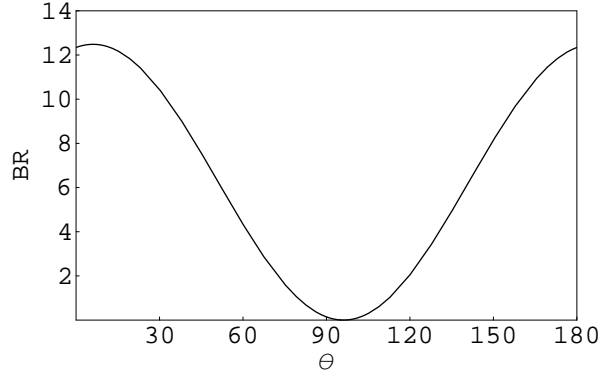}}
 \caption{Same as Figure-1, for the  $B^0 \to   f_0 K^0 $
process.  }
  \end{figure}

\section{CP violation parameters}

Here, we briefly present the basic and well known formula for the CP violating
parameters. Let us first consider the process $B^+ \to f_0 K^+$,
which has only direct CP violation. The amplitude
for this process can be symbolically  written as
\bea
{\cal A}(B^+ \to f_0 K^+) &=& \lambda_u^* |A_u| e^{i \delta_u}
+\lambda_t^* |A_t| e^{i \delta_t}\;,\nn\\
{\cal A}(B^- \to f_0 K^-) &=& \lambda_u |A_u| e^{i \delta_u}
+\lambda_t |A_t| e^{i \delta_t}\;,
\label{dcp}
\eea
where $\lambda_q= V_{qb}V_{qs}^*$ with $(q=u,t)$ denote
the product of CKM matrix elements which contain the
weak phase information. It should be noted that the 
weak phase of $\lambda_u^*$ is
Arg$(V_{ub}^*V_{us})=\gamma$ and that of $\lambda_t^*$ is 
Arg$(V_{tb}^*V_{ts})=\pi$. $A_u$ and $A_t$ denote
the contributions arising from the current operators
proportional to $\lambda_u$
and $\lambda_t$ respectively and the corresponding strong phases 
are taken as $\delta_u$ and $\delta_t$.

For the charged $ B^\pm \to f_0 K^\pm$ decays 
the CP violating rate asymmetry in the partial rates is defined as follows :
\bea
 A_{\rm CP}&=& \frac{\Gamma(B^+ \to f_0 K^+)-\Gamma(B^- \to f_0 K^-)}
{\Gamma(B^+ \to f_0 K^+)+\Gamma(B^- \to f_0 K^-)}\nn\\
&=& \frac{2 r\sin \gamma \sin (\delta_u -\delta_t)}
{1+r^2
-2 r\cos \gamma \cos (\delta_u -\delta_t)}\;,\label{dcp1}
\eea
where $r=|\lambda_u A_u/\lambda_t A_t|$. Thus to obtain significant 
direct CP asymmetry, one requires the two interfering
amplitudes to be of same order and their relative strong phase should be
significantly large (i.e., close to $\pi/2$). However, in the SM, the ratio 
of the CKM matrix elements of the two terms in Eq. (\ref{dcp}) 
can be given (in the Wolfenstein
parametrization) as $|\lambda_u/\lambda_t|\simeq \lambda^2 \sqrt{\rho^2+\eta^2}
\simeq 2 \% $. Therefore, the first amplitude will be highly suppressed 
with respect to the second unless $A_u >> A_t$.   
Hence, the naive expectation is that the direct
CP violation in the SM in this mode will be negligibly small.
Using the generalized factorization approach, we 
find $r=0.15$, $\delta_t - \delta_u\sim 7^\circ$ and the direct CP 
violation in the mode $B^+ \to f_0 K^+ $ as of $\sim(-4 \%)$.
This in turn makes the mode interesting to look for the NP in terms of
large direct CP asymmetry.   

In the presence of new physics the amplitude can be written as
\be
{\cal A}(B^+ \to f_0 K^+)=A_{SM}+A_{NP}= A_{SM}\left [ 1+ r_{NP}~ e^{i
\phi_{NP}} \right ]\;,\label{br11}
\ee
where $r_{NP}=|A_{NP}/A_{SM}|$, ($A_{SM}$
and $A_{NP}$ correspond to the SM and NP contributions to the $B^+
\to f_0 K^+$ decay amplitude, respectively) and $\phi_{NP}={\rm
arg}(A_{NP}/A_{SM} )$, which contains both strong and weak phase
components.
The branching ratio for the $B^+ \to f_0 K^+$ decay process can be given as
\be
BR(B^+ \to f_0 K^+)= BR^{SM}\left ( 1+r_{NP}^2 +2 r_{NP} \cos
\phi_{NP} \right )\;,
\ee
where $ BR^{SM}$ represents the corresponding standard model value.

Now, we will present the basic formula of CP asymmetry
parameters in the presence of new physics. Due to the contributions 
from new physics, these
parameters deviate substantially from their standard model values.
To find out the CP asymmetry, it is necessary to represent explicitly 
the strong and
weak phases of the SM as well as of NP amplitudes. Although, 
it is expected that
the SM amplitude $\lambda_u A_u$ is highly suppressed with 
respect to its
$\lambda_t A_t$ counterpart, for completeness we will keep this term
for the evaluation of ${A}_{\rm CP}$. We denote the NP contribution
to the decay amplitude as $A_{NP}=|A_{NP}| e^{i(\delta_n+\theta_n)}$,
where $\delta_n$ and $\theta_n$ denote the strong and weak phases of the NP
amplitude, respectively. Thus, in the presence of NP, we can explicitly 
write the decay amplitude for
$B^+ \to f_0 K^+$ mode as
\be
{\cal A}(B^+ \to f_0 K^+) =\lambda_u^* |A_u| e^{i \delta_u}
+\lambda_t^* |A_t| e^{i \delta_t}+|A_{NP}| e^{i(\delta_n+\theta_n)}
\;.\label{al}
\ee
The amplitude for $B^- \to f_0 K^-$ mode is obtained by changing the
sign of the weak phases of the amplitude (\ref{al}). Thus,
the CP asymmetry parameter  is given as
\be
{A}_{\rm CP} =\frac{2\biggr(r
\sin \gamma \sin \delta_{ut}+r_N
\sin \theta_n \sin \delta_{nt}-r r_N
\sin (\gamma-\theta_n) \sin \delta_{un}
\biggr)}
{|{\cal A}|^2-
2\biggr(r
\cos \gamma \cos \delta_{ut}+
r_N
\cos \theta_n \cos \delta_{nt}-r r_N
\cos (\gamma-\theta_n) \cos \delta_{un}\biggr)},
\ee
where $|{\cal A}|^2=1+r^2+r_N^2$, $r_N= |A_{NP}/\lambda_t A_t|$
and $\delta_{ij}=\delta_i-\delta_j $ are the relative strong phases
between different amplitudes.

Now we consider the CP violation parameters in the neutral $B$ meson
decays, which has both direct and mixing-induced components.
Let us consider the $B^0$ and ${\bar B}^0$ decay into a
CP eigenstate $f_{CP}$ (we consider $f_{CP}=f_0 K_S  $ with CP
eigenvalue +1).

The time dependent CP asymmetry for $B \to f_0 K_S$
can be described as \cite{nir}
\bea
{A}_{f_0 K_S}(t) &=& \frac{\Gamma (B^0(t) \to f_0 K_S)-
\Gamma (\bar B^0(t) \to f_0 K_S)}{\Gamma (B^0(t) \to f_0 K_S)+
\Gamma (\bar B^0(t) \to f_0 K_S)}\nn\\
&=& C_{f_0 K_S}\cos (\Delta M_{B_d} t)- S_{f_0 K_S}
\sin (\Delta M_{B_d} t)\;,
\eea
where we identify
\be
C_{f_0 K_S}=\frac{1-|\lambda|^2}{1+|\lambda|^2},~~~~
    S_{f_0 K_S}=\frac{2 {\rm Im}(\lambda)}{1+|\lambda|^2}\;,
\ee
as the direct and the mixing-induced CP asymmetries.
The parameter $\lambda$ corresponds to
\be
\lambda=\frac{q}{p}\frac{
{\cal A}(\bar B^0 \to f_0 K_S)}{{\cal A}(B^0 \to f_0 K_S)}\;,
\ee
where, $q$ and $p$ are the mixing parameters and are
represented by the CKM
elements in the standard model as
\be
\frac{q}{p}=\frac{V_{tb}^* V_{td}}{V_{tb} V_{td}^*}\sim 
{\rm exp}(-2 i \beta)\;.
\ee
Now the amplitude for $ \bar B^0 \to f_0 K_S$ 
can be symbolically written  as 
\be
{\cal A}(\bar B^0 \to f_0 K_S)=\lambda_t A_t\;,
\ee
where $\lambda_t= V_{tb}V_{ts}^*$, which is real in the SM.  Thus, 
the mixing induced CP asymmetry 
is given as,
$S_{f_0 K_S}=-\sin 2 \beta $,
same in magnitude as the one for $B \to \psi K_S$, but 
with opposite
 sign and the direct CP
asymmetry turns out to be identically zero. 

However, the decay amplitude
also receives some contribution from the internal {\it up} and 
{\it charm} quarks in the loop. Therefore, the CP violating
parameters may deviate from their expected values. 
Now including the effects of $u, c,t$
quarks in the loop and using CKM unitarity 
($\lambda_u+\lambda_c +\lambda_t = 0$), one can write the decay
amplitude as
\bea
{\cal A}( B^0 \to f_0 K^0) = \lambda_u^* A_{u} + \lambda_c^* A_{c}
=\lambda_c^* A_{c}\left [1+r' e^{i(\delta' +\gamma)}\right ]\;,
\eea  
where the amplitude  $A_u$ contains contributions from $u$ and
$t$ quarks in the loop (i.e., $A_u=P_u-P_t$, where $P_{u,c,t}$
are the penguin amplitudes corresponding to $u,c,t$ quark exchange
in the loop)
and same argument holds for $A_c$. The parameter $r'$ is the
ratio of the two amplitudes, i.e., 
$r'=|\lambda_u A_u/\lambda_c A_c|$,  $\delta'= \delta_{u}-\delta_{c}=
{\rm Arg}(A_u/A_c)$ 
is the relative strong 
phase  between them and  $\gamma$ is the weak phase.
The explicit expressions for these amplitudes 
(in units of $-G_F/\sqrt 2$) are given as 
\bea
A_q &= &\Big[a_4^q-\frac{a_{10}^q}{2}
-r_{\chi_1}(a_6^q-\frac{a_8^q}{2})\Big] X 
+(2 a_6^q -a_8^q) Y\nn\\
&-&\Big[ a_4^q-\frac{a_{10}^q}{2}
-\frac{(2a_6^q-a_8^q) m_B^2}{(m_b+m_d)(m_s+m_d)}\Big] Z \;,
\eea
with $q=u$ and $c$, $Y$ and $Z$ are given in Eq. (12). 
Thus, one obtains the CP asymmetries as
\bea
S_{f_0 K_S} &=& -\frac{\sin 2 \beta + 2r' \cos \delta^\prime 
\sin(2 \beta+\gamma)
+r'^2 \sin (2\beta +2\gamma)}{1+r'^2+2 r' \cos \delta^\prime 
\cos \gamma}\;,\nn\\
C_{f_0 K_S}&=& \frac{-2r' \sin \delta^\prime \sin \gamma}
{1+r'^2+2r' \cos \delta^\prime
\cos \gamma}\;.\label{cpa}
\eea
In order to know the precise value of the CP
violating asymmetries one should know the values of $r'$ and 
$\delta^\prime$. Using the QCD coefficients from \cite{cheng}
we obtain $
r'=0.02$, $\delta^\prime=12^\circ$ and hence the CP asymmetries as
\bea
S_{f_0 K_S}=-0.672\;~~~~{\rm and}~~~~C_{f_0 K_S}=-0.007\;,
\eea
which are in accordance with the results of top quark
dominance in the penguin loop. Therefore, here onwards we will consider
the SM amplitude for the
$B^0 \to f_0 K_S$ process to be dominated by the top quark 
penguin. 


New physics could in principle contribute to both mixing and decay
amplitudes. The new physics contribution to mixing is universal
while it is non-universal and process dependent in the decay
amplitudes. As the NP contributions to mixing phenomena is
universal, it will still set $ S_{\psi K_S}=-S_{f_0 K_S}$.
Therefore, to explain the  deviation between $
(S_{\psi K_S}$ and $(\sin 2 \beta)_{f_0 K_S}=-S_{f_0 K_S})$, 
here we explore the NP effects only
in the decay amplitudes. Thus, including the NP contributions, we
can write the decay amplitude for $ B \to f_0 K$ process as
\be
{\cal A}(B^0 \to f_0 K^0)=A_{SM}+A_{NP}= \lambda_t^* A_t \left [ 1- r_{N}~ 
e^{i(\delta_{nt}+\theta_n)} \right ]\;,
\ee
where $r_{N}=|A_{NP}/\lambda_t A_{t}|$, $\delta_{nt}$ and $\theta_{n}$ are 
the relative strong and weak
phases between the new physics contributions to the decay
amplitude and that of the SM part. The negative sign
before $r_N$ in Eq. (35) arises because the weak phase 
$\pi$ of $\lambda_t^*$ has been factored out. Thus, one can then obtain the 
expressions for
the CP asymmetries as
\be S^{NP} =-\frac{\sin 2 \beta- 2
r_{N} \cos \delta_{nt} \sin(2\beta +\theta_{n})+r_{N}^2 \sin (2
\beta+2 \theta_{n})}{ 1+r_{N}^2 -2 r_{N} \cos \delta_{nt}\cos
\theta_{n}}\;,\label{eq:eq1}
 \ee
 and
\be
 C^{NP} =\frac{2
r_{N} \sin \delta_{nt} \sin \theta_{n}}{ 1+r_{N}^2 -2 r_{N}
\cos \delta_{nt}\cos \theta_{n}}\;.\label{eq:eq2}
 \ee

Having obtained the CP asymmetry parameters, in 
the presence of new physics, we now proceed to evaluate the same 
in the R-parity violating supersymmetric model.

\section{ Contribution from R-parity violating supersymmetric model}

We now analyze the decay modes in the minimal supersymmetric model with
R-parity violation.
In the supersymmetric models there may be interactions which
violate the baryon number $B$ and the lepton number $L$
generically. The simultaneous presence of both $L$ and $B$ number
violating operators induce rapid proton decay, which may contradict
strict experimental bound. In order to keep the proton lifetime
within experimental limit, one needs to impose additional symmetry
beyond the SM gauge symmetry to force the unwanted baryon and lepton
number violating interactions to vanish. In most cases this has
been done by imposing a discrete symmetry, called R-parity defined
as, $R_p=(-1)^{(3B+L+2S)}$, which is +1 for all particles
and $-1$ for all superparticles.
This symmetry not only forbids rapid proton decay, but also 
prevents single creation and annihilation of superparticles.
However, this
symmetry is ad-hoc in nature. There is no theoretical arguments in
support of this discrete symmetry. Hence, it is interesting to see
the phenomenological consequences of the breaking of R-parity in
such a way that either $B$ or $L$ number is violated, both 
not simultaneously violated, thus avoiding rapid proton decay.
Extensive studies have been done to look for the direct as well as
indirect evidence of R-parity violation from different processes
and to put constraints on various R-parity violating couplings.

For our purpose, we will consider  the
Lepton number violating
super-potential with only $\lambda^\prime$ couplings, which
is given as
\begin{equation}
W_{\not\!{L}} =\lambda_{ijk}^\prime L_i Q_j D_k^c \;,\label{eq:eqn10}
\end{equation}
where, $i, j, k$ are generation indices, $L_i$ and $Q_j$ are
$SU(2)$ doublet for lepton and quark superfields and $D_k^c$
is the down type quark singlet superfield. 

Thus the effective Hamiltonian for charmless hadronic 
$B$ decays can be given as
\cite{dutta}
\bea
{\cal H}_{eff}^{\lp}  &= & d_{jkn}^R[\bar d_{n \alpha}\gamma_L^\mu d_{j \beta}
\bar d_{k \beta}\gamma_{\mu R} b_{ \alpha}]+
d_{jkn}^L[\bar d_{n \alpha}\gamma_L^\mu b_{ \beta}
\bar d_{k \beta}\gamma_{\mu R} d_{j \alpha}]\nn\\
&+& u_{jkn}^R[\bar u_{k \alpha}\gamma_L^\mu u_{j \beta}
\bar d_{n \beta}\gamma_{\mu R} b_{ \alpha}]\;,
\eea
where $\alpha$, $\beta$ are the color indices, $\gamma_{R,L}^\mu
=\gamma^\mu(1 \pm \gamma_5)$ and
\be
d_{jkn}^R= \sum_{i=1}^3 \frac{\lp_{ijk} \lambda_{in3}^{\prime *}}{8 m_{\tilde
\nu_{Li}}^2}\;,~~~~~
d_{jkn}^L= \sum_{i=1}^3 \frac{\lp_{i3k} \lambda_{inj}^{\prime *}}{8 m_{\tilde
\nu_{Li}}^2}\;,~~~~
u_{jkn}^R= \sum_{i=1}^3 \frac{\lp_{ijn} \lambda_{ik3}^{\prime *}}{8 m_{\tilde
e_{Li}}^2}\;.
\ee

Thus one can write the transition amplitudes as
\bea
{\cal A}^{\lp}(B^+ \to f_0 K^+) &=&-2(d_{222}^L+d_{222}^R) Y+
u_{112}^R r_\chi X-(d_{112}^R+d_{121}^L)2 Y_d\;, \nn\\
{\cal A}^{\lp}(B^0 \to f_0 K^0) &=&-2(d_{222}^L+d_{222}^R) Y+
(d_{121}^R+d_{112}^L )r_{\chi_1} X\nn\\
&+&(d_{112}^R+d_{121}^L ) 
\left ( \frac{X}{N}-2Y_d \right )\;,
\eea
where $Y_d$ is the value of Y with $\bar f_{f_0}^s$ replaced by
 $\bar f_{f_0}^d$.

Following the standard practice,  we shall assume that
the RPV couplings are hierarchical, i.e., only one
combination of the coupling is numerically
significant. Furthermore, we also
assume that both the transitions $B^{+,0} \to f_0 K^{+,0}$
receive dominant contribution from the quark level transition 
$b \to s \bar s s$, and hence we consider only $d_{222}^L$ coupling to be 
nonzero. As discussed in Ref. \cite{kundu}, we will also discard 
the $d_{222}^R$ 
coupling in our analysis, as it is related to $u_{222}^R$ by SU(2)
isospin symmetry and its effect in the mode $B \to J/\psi K_S$ 
is found to be negligibly small. Thus, with these approximations 
the transition amplitudes 
for both the processes can be given as
\bea
{\cal A}^{\lp}(B \to f_0 K) =
-\frac{1}{8 m^2_{\tilde \nu_{Li}}} 
\Big(\lambda_{i32}^{\prime } \lambda_{i22}^{\prime *}
\Big)
2 m_{f_0} \bar f_{f_0}^s \frac{m_B^2-m_K^2}{m_b-m_s}
F_0^{BK}(q^2)\;,
\eea
where the summation over $i=1,2,3$ is implied.
Now considering the values of R-parity couplings 
as 
\be
\lambda_{i32}^{\prime } \lambda_{i22}^{\prime *}
=R
e^{i \theta_n}\;,
\ee
where  $R = |\lambda_{i32}^{\prime } \lambda_{i22}^{\prime *}| $
and $\theta_n$ is the new weak phase  with range
$-\pi \leq \theta_n \leq  \pi$. It should be noted that
since the dominant SM amplitude (i.e., the t-quark
dominated penguin amplitude $A_t$) contains the weak phase
$\pi$, we vary the weak phase $\theta_n$ between [$-\pi,\pi$],
so that the NP amplitude will interfere constructively with 
the SM amplitude 
when the relative weak phase between them is zero.
To see the effect of R-parity violation in the decay modes
$ B \to f_0(980) K$, it is essential to know the value
of the RPV couplings ($R$). We first present a crude estimation of $R$
by assuming that R-parity will explain the observed discrepancy
between the observed and SM predicted branching ratios for
$B \to f_0 K $ modes.
We will further assume that the new physics amplitude will
interfere constructively with the standard model amplitude
(i.e., $\phi_{NP}=0$ in Eq. (22)), so that
one can obtain a lower bound on $r_{NP}$ from Eqn. (23). Now using
the values of the experimental branching ratios from Eq. (4) and 
the corresponding SM values from (19), we obtain
the lower bound as $r_{NP}\geq 0.6$.
This, in turn with Eqns. (11), (14) and (42) gives
\be
R \geq 1 \times 10^{-3}\;,
\ee
for $ m_{\tilde \nu_{Li}}=100$ GeV.
Recently, in Ref. \cite{yadong} it has been shown that
the branching ratio and the polarization anomaly in $B \to \phi K^*$ 
modes can be resolved in the R-parity violating supersymmetric model
for a very narrow interval in the parameter space as
 $|\lambda_{i32}^{\prime } 
\lambda_{i22}^{\prime *}|/m_{\tilde \nu_{Li}}^2\in [1.5 \times 10^{-3} ,
2.1 \times 10^{-3} ]$, for the sneutrino  mass scale 
$ m_{\tilde \nu_{Li}}=100$ GeV.
Therefore, in this analysis we consider the lowest value for
$R$  i.e., $ 1.5 \times 10^{-3} $ from the above allowed range, which
also satisfies the constraint (44).
Using this value we obtain the ratios of RPV to SM amplitude, as
defined in section III, as
\bea
&&r_{NP}=0.81\;, ~~~~r_N=0.87   ~~~({\rm for}~~B^+ \to f_0 K^+)\;,\nn\\ 
&&r_{N}=0.92\;,~~~~~ ({\rm for}~~B^0 \to f_0 K^0)\;.
\eea
Therefore, the upper limits in
the branching ratios (for $\phi_{NP}=0$ in Eq.(23)) 
in the RPV model are found to be
\bea
BR(B^+ \to f_0(980) K^+ )& \leq & 21.6
\times 10^{-6}\;, \nn\\
BR(B^0 \to f_0(980) K^0 ) & \leq & 17.4
\times 10^{-6}\;.
\eea
Thus one can see that the observed branching ratios (4) can be accommodated
in the RPV model.
 
Now assuming the strong phase difference $\delta_{nt}$ to be small 
(e.g., $\sim 10^\circ$),
direct CP violation for $B^+ \to f_0 K^+ $ process and the mixing induced CP
violating parameter from $ B^0 \to f_0K^0$ are shown in figures - 3 and 4.
Thus, as seen from the figures, the observed 
$(\sin 2 \beta)_{f_0 K_S}=-S_{f_0 K_S}=0.51\pm
0.19$ can be explained 
in the RPV model and large direct CP violation (upto 80 \%) in $B^+ 
\to f_0 K^+ $ mode could be obtainable in this model. However, 
there  is no obvious reason why the strong phase difference  
$\delta_{nt}$ could
be small. To see the impact of the strong phase  we vary
it between the range of $-\pi$ and $ \pi$ and plot the correlation between
direct and mixing induced CP asymmetries for $B^0 \to f_0 K_S $, for two
representative values of weak phase $\theta_n=\pi/2$ and $\pi/4$,
in figure-5. From the figure, it is seen that R-parity
violating supersymmetric model can accommodate large CP violation 
in the $B^0 \to f_0 K_S$ decay mode.

\begin{figure}[htb]
   \centerline{\epsfysize 2.0 truein \epsfbox{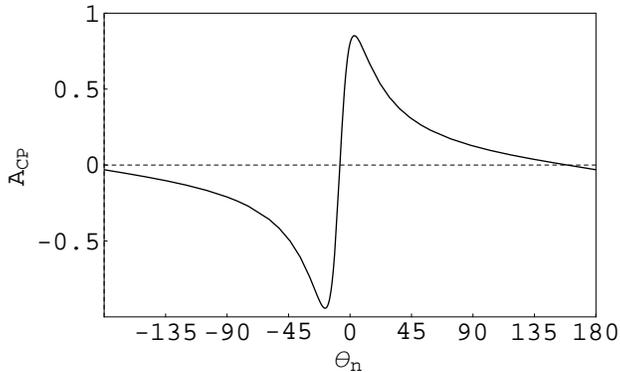}}
 \caption{
 Direct CP violation  for the process  
$B^+ \to   f_0(980) K^+ $  versus the new weak phase 
 $\theta_n$ in degrees.}
  \end{figure}
\begin{figure}[htb]
   \centerline{\epsfysize 2.0 truein \epsfbox{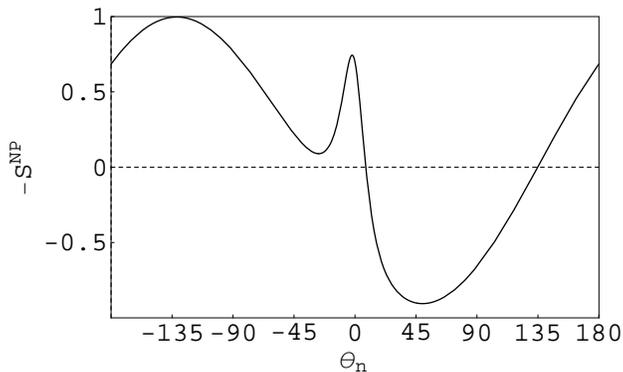}}
 \caption{
 Mixing induced CP violation  for the process  
$B^0 \to   f_0(980) K^0 $  versus the new weak phase 
 $\theta_n$ in degrees.}
  \end{figure}
\begin{figure}[htb]
   \centerline{\epsfysize 2.25 truein \epsfbox{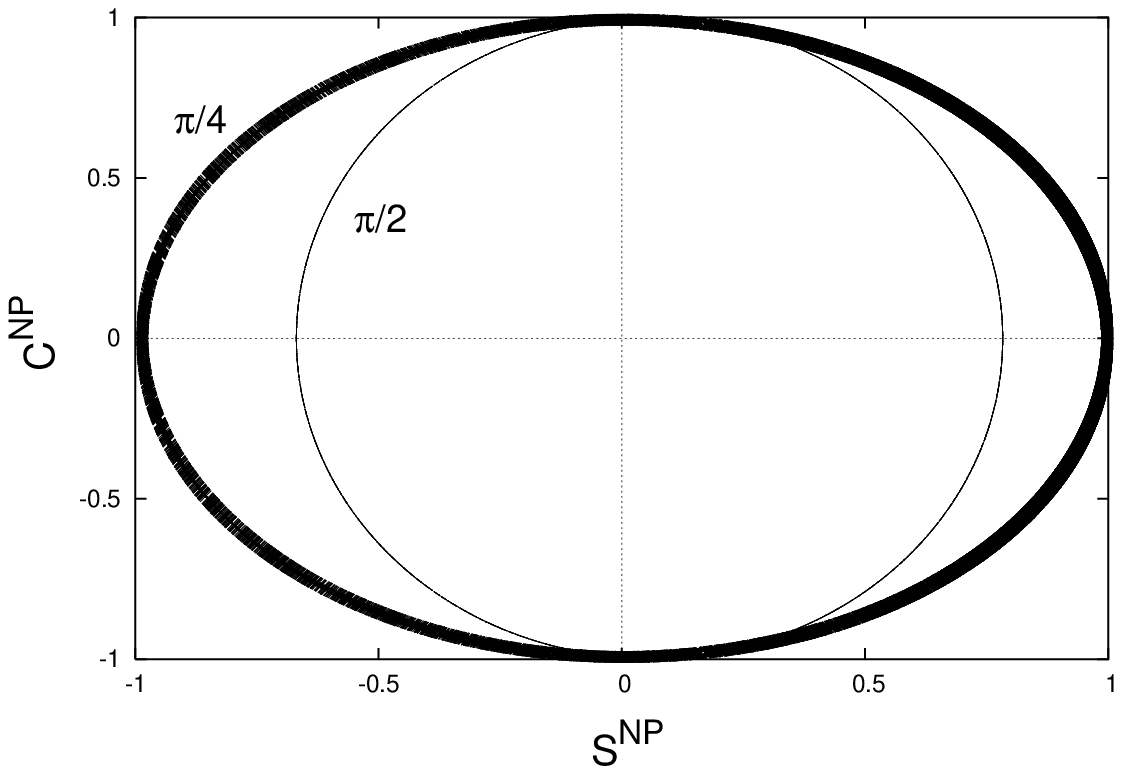}}
 \caption{
 The correlation plot between $S^{NP}$ and $C^{NP}$  for the process  
$B^0 \to   f_0(980) K_S $  in the RPV model for two representative 
values of weak phases ($\theta_n=\pi/2,~\pi/4$), where we have used 
$r_N=0.92$,  and varied
the strong phase $\delta_{nt} $ between $-\pi$ and $ \pi$.
}
  \end{figure}
\section{conclusion}

In this paper we have studied the rare decay modes $B \to f_0(980) K$,
involving a scalar and a pseudoscalar meson in the final state.
Since the structure of the $f_0$ meson is not well established
till now, we consider it as a $ q \bar q $ state, comprising of both
$s \bar s$ and $(u \bar u+ d \bar d)/\sqrt 2 $ components with a
mixing angle of $138^\circ$, which appears to be the most preferable one. 
Using the
generalized factorization approach, we found that the branching ratios
in the standard model are below the current experimental values, as was
obtained in previous studies using different approaches. The 
average value of the observed
 mixing induced CP asymmetry, i.e., $(\sin 2 \beta)_{f_0 K_S}=-
S_{f_0 K_S}$, also has about 
one sigma deviation
from that of $S_{J/\psi K}$. To explain the observed discrepancy in
the branching ratios and CP asymmetry parameter, we considered the R-parity
violating model. Since these processes receive dominant contribution
from $b \to s \bar s s$ loop induced penguins, we assumed that
the new physics parameters will affect such transitions strongly.
We found that the R-parity violating model can explain the observed
discrepancy in the branching ratios and the CP violation 
parameter $-S_{f_0K_S}$. It can accommodate large CP violation even for
small relative strong phase between SM and RPV amplitudes. 
In this analysis, we have considered a representative value for the
RPV coupling $|\lambda_{i32}^{\prime } \lambda_{i22}^{\prime *}|$. But,
it should be noted that using the data on the branching ratios and CP 
asymmetries of the processes, which have dominant $b \to s \bar s s$
quark level transitions, it would be possible to obtain the allowed
parameter space  for the magnitudes and phases of the RPV couplings.
If, in
future, the
$q \bar q$ structure for $f_0$ is established then these modes could also
play an important role to look for new physics beyond the standard model or
else, at least, it will certainly enrich our understanding regarding the
nature of the light scalar mesons.

\acknowledgments

The work of RM was partly supported by Department of Science and Technology,
Government of India,
through grant Nos. SR/FTP/PS-50/2001 and
SR/S2/HEP-04/2005. BM would like to thank Council of Scientific
and Industrial Research, Government of India, for financial support. 


\end{document}